\documentclass[conference]{IEEEtran}
\IEEEoverridecommandlockouts
\usepackage{cite}
\usepackage{amsmath,amssymb,amsfonts}
\usepackage[ruled]{algorithm2e}
\usepackage{graphicx}
\usepackage{textcomp}
\usepackage{siunitx}
\usepackage{xcolor}
\usepackage{breqn} 
\usepackage{subfig} 
\usepackage{tabularx} 
\usepackage{booktabs}

\newcolumntype{P}[1]{>{\centering\arraybackslash}p{#1}} 
\newtheorem{remark}{Remark}
\def\BibTeX{{\rm B\kern-.05em{\sc i\kern-.025em b}\kern-.08em
    T\kern-.1667em\lower.7ex\hbox{E}\kern-.125emX}}
\begin{document}

\title{Decentralized Linear MMSE Equalizer Under Colored Noise for Massive MIMO Systems
 }\author{aa}
\author{\IEEEauthorblockN{Xiaotong Zhao$^{1}$,
Xin Guan$^{1}$,
Mian Li$^{2,3}$, and
Qingjiang Shi$^{1,3}$
}
\IEEEauthorblockA{$^{1}$School of Software Engineering, Tongji University, Shanghai, China}
\IEEEauthorblockA{$^{2}$School of Elec. Info. and Comm., Huazhong University of Science and Technology, Wuhan, China}
\IEEEauthorblockA{$^{3}$Shenzhen Research Institute of Big Data, Shenzhen, China}
}

\maketitle

\begin{abstract}
  Conventional uplink equalization in massive MIMO systems relies on a centralized baseband processing architecture. However, as the number of base station antenna increases, centralized baseband processing architectures encounter two bottlenecks, i.e., the tremendous data interconnection and the high-dimensional computation. To tackle these obstacles, decentralized baseband processing was proposed for uplink equalization, but only applicable to the scenarios with unpractical white Gaussian noise assumption. This paper presents an uplink linear minimum mean-square error (L-MMSE) equalization method in the daisy chain decentralized baseband processing architecture under colored noise assumption. 
The optimized L-MMSE equalizer is derived by exploiting the block coordinate descent method, which shows near-optimal performance both in theoretical and simulation while significantly mitigating the bottlenecks.
\end{abstract}


\section{Introduction}
Massive multiple-input multiple-output (MIMO) is considered as one of the key enabling technologies for future B5G (e.g., 5.5G recently proposed by Huawei \cite{5-5G}) and 6G systems due to its ability to enhance both the spectrum and power efficiency \cite{mimo1,mimo2,mimo3}. Benefit from the equipment of large-scale antenna arrays, a base station (BS) in a massive MIMO system can simultaneously serve large numbers of user equipments (UEs)  given the same time-frequency resource.

Conventional uplink equalization/data detection schemes in massive MIMO systems, such as zero-forcing (ZF) or linear minimum mean-square error (L-MMSE) equalization methods, rely on a centralized baseband processing architecture \cite{li2017baseband}. However, with the rapid growth of the number of antennas at the BS, such centralized methods suffer from tremendous raw data interconnection between the baseband processing units and the radio frequency chains \cite{li2017baseband,Rodriguez2020}. For a typical 256-antenna BS with 80\si{MHz} bandwidth and 12-bit digital-to-analog converters (DACs), the raw baseband data interconnection throughput reaches 1Tbps, which greatly exceeds the capacity of the existing data interconnection standards such as the enhanced common public radio interface (eCPRI) \cite{eCPRI}. Meanwhile, as the number of antennas increases, the equalization and other baseband processing procedures at the BS may encounter complicated matrix calculation (e.g., the inverse of an extremely high dimensional matrix),  
which leads to unbearable high computational complexity \cite{li2017baseband}.

\begin{figure*}[!tb]
    \centering
    \subfloat[Architecture of a uni-directional daisy chain with loop]{\includegraphics[width=0.4\textwidth]{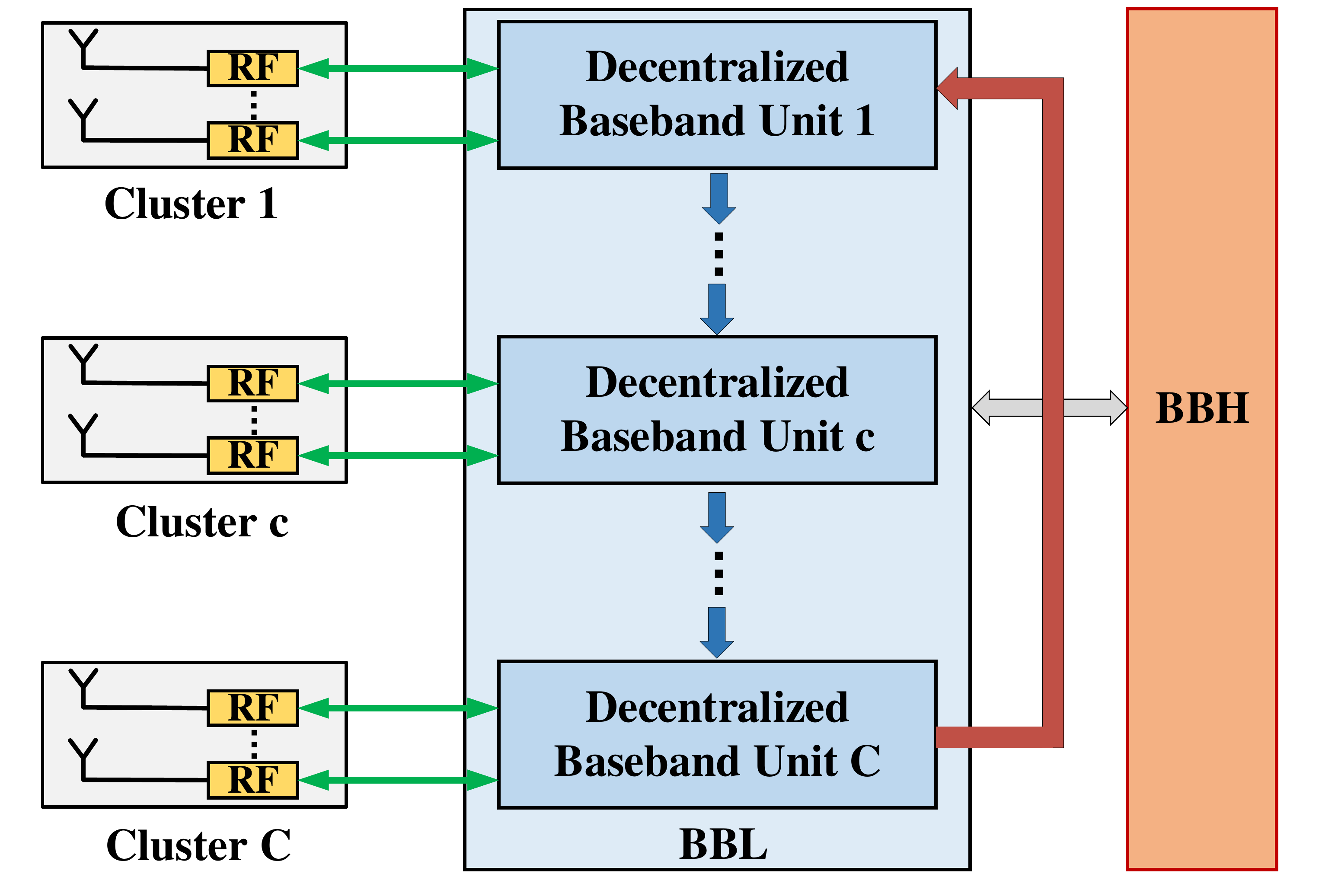}
        \label{fig_second_case}}
    \hfil
    \subfloat[Architecture of a bi-directional daisy chain]{\includegraphics[width=0.4\textwidth]{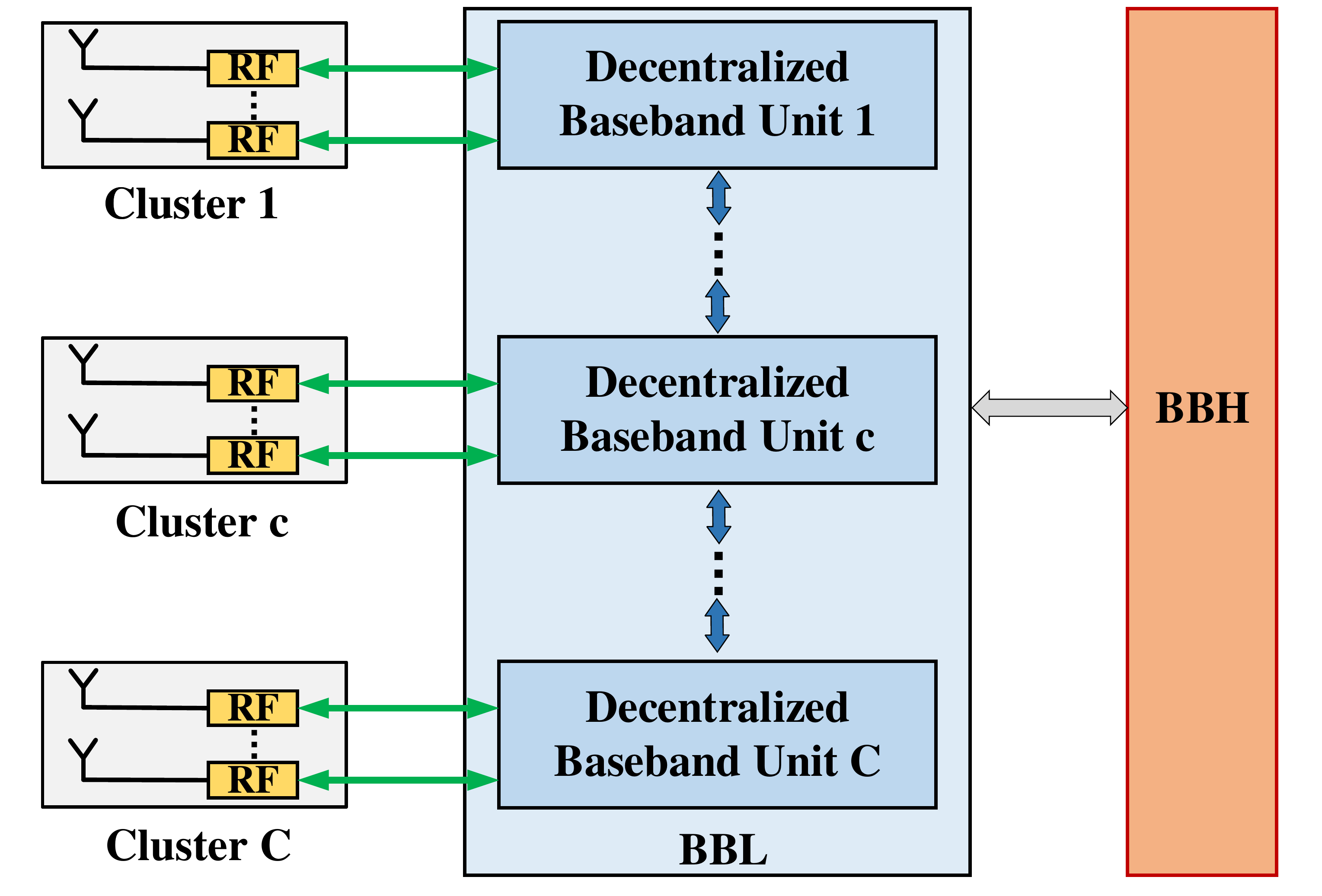}
        \label{fig_first_case}}
    \caption{The $M$ BS antenna elements are divided into $C$ clusters, and each is associated with local radio-frequency (RF) and a computing fabric, which is called a decentralized baseband unit (DBU). Each DBU independently performs decentralized data detection. (a) The DBUs are connected by uni-directional links, and there is an extra link connecting the last and the first DBUs. (b) The DBUs are connected by bi-directional links. Note that architecture (b) has fewer interfaces, leading to lower connection costs and higher latency than architecture (a). The proposed algorithm aims at architecture (a), but it can also be applied to architecture (b) after generalization.}
    \label{fig_architecture}
\end{figure*}

To overcome the two bottlenecks in conventional centralized baseband processing architectures, i.e., the tremendous data interconnection and the high-dimensional computation, a prevailing solution is to exploit decentralized baseband processing (DBP) \cite{li2017baseband,Rodriguez2020,li2019cd,Jeon2019feedforward,Sanchez2019discuss,li2019design}. As shown in Fig.~\ref{fig_architecture}, in DBP architecture, the original centralized baseband processing unit is replaced by multiple distributed local baseband units (DBUs). Moreover, the BS antennas are also partitioned into multiple independent clusters, such that each cluster is connected to a DBU. As a result, each DBU can only acquire local information (e.g., channel state information, receive signals, noise samples) of the corresponding antenna cluster. Various works have investigated effective transmission designs in the DBP architecture.

To the best of our knowledge, previous works \cite{li2017baseband,Rodriguez2020,li2019cd,Jeon2019feedforward,Sanchez2019discuss} only consider DBP designs with the ideal additive white Gaussian noise (AWGN), i.e., the correlation matrix is assumed to be diagonal. With such a nice property, the correlation matrix can be naturally decomposed into multiple diagonal submatrices, which perfectly suits the implementation of DBP architecture. Unfortunately, this no longer holds when considering more practical colored noise\footnote{The colored noise may arise when inference from other unwanted users exists during the noise estimation process.} with a non-diagonal correlation matrix. Moreover, in a DBP architecture, each DBU only has noise samples with respect to the corresponding antenna cluster. As a result, computing the non-diagonal correlation matrix of colored noise requires collecting noise samples from all antenna clusters. This suffers from prohibitively high data interconnection and computational complexity since the number of noise samples is proportional to the number of antennas at the BS (which may be extremely high). Therefore, DBP designs with both low data interconnection and low computational complexity under colored noise assumption remain a challenge. 

This paper investigates the uplink L-MMSE equalization in a DBP architecture of a massive MIMO system with the colored noise assumption, where the correlation matrix of the colored noise is estimated by averaging multiple noise samples. Rather than directly obtain the estimated symbol, we focus on designing an L-MMSE equalization matrix in a decentralized daisy chain architecture. The considered L-MMSE equalization in DBP architecture is formulated as a convex quadratic optimization problem, which can be efficiently solved by exploiting the block coordinate descent (BCD) method \cite{2016Bertsekas}. With the proposed method, the dimension of interconnected data is significantly reduced, only depending on the number of UEs. Theoretical analysis and simulation results demonstrate that the proposed method can achieve a performance close to the
centralized L-MMSE equalization with much lower computational complexity.
 
\section{DECENTRALIZED EQUALIZATION ARCHITECTURES}
In this section, we first present the uplink massive MIMO system model and review some basic concepts of the L-MMSE equalization, then introduce the decentralized uplink architecture. 
\subsection{System Model and MMSE Equalization}
Consider a massive MIMO system with $K$ single-antenna target UEs transmitting data to a BS equipped with $M$ antenna elements, where $M\gg K$. The BS received signal vector $\mathbf{y} \in \mathbb{C}^{M}$ can be represented by
\begin{equation}\label{cen-model}
    \mathbf{y}=\mathbf{H}\mathbf{s}+\mathbf{n} ,
\end{equation}
where $\mathbf{H} \in \mathbb{C}^{M \times K}$ represents the channel matrix, $\mathbf{s} \in \mathcal{D}^{K}$ denotes the transmitted user data symbol vector with $\mathcal{D}$ representing the constellation set for modulation scheme (e.g., 16-QAM), $\mathbf{n} \sim \mathcal{CN}(\mathbf{0},\mathbf{R})$ is a random vector that models the $M$ dimensional receiver noise with a \emph{non-diagonal} correlation matrix $\mathbf{R}=\mathbb{E}(\mathbf{n}\mathbf{n}^{H})$, which significantly differs from conventional assumptions of AWGN. The non-diagonal correlation matrix assumption is reasonable since 
the noise consists of thermal noise and interference signals of UEs coming from neighboring cells other than straight  AWGN. In other words, the noise in the considered massive MIMO system is assumed to be correlated (colored). In practice, the correlation matrix  $\mathbf{R}$ can be only estimated by averaging the noise samples in $N$ pilot resource elements (REs): 
\begin{equation}\label{equation:estimate_R}
    \hat{\mathbf{R}}=\frac{1}{N}\sum_{i=1}^{N}\mathbf{n}^i(\mathbf{n}^i)^H,
\end{equation}
where $N$ typically equals $96$ or $192$, $\mathbf{n}^i \in \mathbb{C}^{M}$ is the noise sample in the $i$-th pilot RE. Both the colored noise assumption and the inaccurate correlation matrix through sampling make the conventional decentralized equalization design more challenging. We will further discuss it in the next subsection.

Note that in this paper, we are dedicated to obtaining an equalization matrix rather than to directly estimate symbols, in order to significantly reduce the computational complexity and the amount of information exchange. In a typical scenario, the channel impulse response is considered to be almost constant across $N_{\text{coh}}$ contiguous symbols, for which the equalization matrix can be reused, thus significantly reducing the computational overhead \cite{Sanchez2019discuss}.

L-MMSE equalization seeks for a linear matrix, which solves the following problem:
\begin{equation}\label{MMSE-model}
    \min _{\mathbf{W}}  \quad \mathbb{E}\left\| \mathbf{W} \mathbf{y}-\mathbf{s} \right\|_2^{2},
\end{equation}
and leads to the well-known L-MMSE\footnote{Note that the transmitted symbol and receiver noise are both under Gaussian assumption. Thus the MMSE equalization is equivalent to the L-MMSE equalization. Consequently, we drop the L in the following.} receiver:
\begin{equation}\label{MMSE-solution}
        \mathbf{W}_{\text{MMSE}} =(\mathbf{H}^H\mathbf{R}^{-1}\mathbf{H}+\frac{1}{E_{s}}\mathbf{I})^{-1}\mathbf{H}^H\mathbf{R}^{-1},
    \end{equation}
where $E_s$ is the expected per-user transmit energy. The final linear estimate $\hat{\mathbf{s}}_{\text{MMSE}}$ is given by applying the obtained MMSE equalizer filter matrix $\mathbf{W}_{\text{MMSE}}$ to the received vector $\mathbf{y}$, i.e. $\hat{\mathbf{s}}_{\text{MMSE}}=\mathbf{W}_{\text{MMSE}}\mathbf{y}.$

\subsection{Decentralized Uplink Architecture}
As illustrated in Fig.~\ref{fig_architecture}(a), in the considered decentralized uplink architecture, the $M$ BS antennas are partitioned into $C$ antenna clusters, where the $c$-th cluster consists of $M_c$ antennas with $M=\sum_{c=1}^{C} M_{c}$. Thus we partition the received vector $\mathbf{y}=[\mathbf{y}_{1}^{T},\mathbf{y}_{2}^{T},\cdots,\mathbf{y}_{C}^{T}]^{T}$, the channel matrix $\mathbf{H}=[\mathbf{H}_{1}^{T},\mathbf{H}_{2}^{T},\cdots,\mathbf{H}_{C}^{T}]^{T}$, and the noise vector $\mathbf{n}=[\mathbf{n}_{1}^{T},\mathbf{n}_{2}^{T},\cdots,\mathbf{n}_{C}^{T}]^{T}$ in equation \eqref{cen-model}. Therefore, the received signal $\mathbf{y}_{c} \in \mathbb{C}^{M_{\mathrm{c}}}$ at the $c$-th cluster can be represented by
\begin{equation}\label{decen-model}
    \mathbf{y}_{c}=\mathbf{H}_{c} \mathbf{s}+\mathbf{n}_{c}, \quad c=1,2, \ldots, C,
\end{equation}
where $\mathbf{H}_{c} \in \mathbb{C}^{M_{c} \times K}$ (a sub-matrix of $\mathbf{H})$ is the local channel matrix, and $\mathbf{n}_{c} \in \mathbb{C}^{M_{c}}$ is the local noise vector at cluster $c$.

Note that all the noise samples $\mathbf{n}^i$ are stored in the decentralized architecture by $\mathbf{n}^i=[(\mathbf{n}_1^i)^{T},(\mathbf{n}_2^i)^{T},\cdots,(\mathbf{n}_C^i)^{T}]^{T}$, thus the noise correlation matrix $\mathbf{R}$ can be regarded as a block matrix with $C \times C$ blocks, where the $(m, n)$-th block is denoted by $\mathbf{R}_{mn}=\mathbb{E}(\mathbf{n}_m\mathbf{n}_n^{H})$. Since accurate estimation of $\mathbf{R}$ is necessary to ensure the good performance of MMSE equalization, and each cluster $c$ can only locally estimate $\mathbf{R}_{cc}$ by $\hat{\mathbf{R}}_{cc}=(1/N)\sum_{i=1}^{N}\mathbf{n}^i_c(\mathbf{n}^i_c)^H$, the critical point lies in how to accurately obtain the off-diagonal blocks (i.e., $\mathbf{R}_{mn}, m \ne n$) of $\mathbf{R}$. Since noise samples are distributed stored in each 
DBU, the direct exchange of noise samples will inevitably induce a large amount of information exchange. Consequently, it is difficult to obtain the MMSE equalization matrix as \eqref{MMSE-solution} in a decentralized manner due to the bandwidth limitation. 

An intuitive way to tackle this is to approximate $\mathbf{R}$ by a block diagonal matrix via setting the off-diagonal blocks to be zero matrix, and denote it by $\mathbf{R}_{\text{block}}=\text{diag} \left(\mathbf{R}_{11},\mathbf{R}_{22},\cdots,\mathbf{R}_{CC}\right)$. In this way, the MMSE equalization matrix in \eqref{MMSE-solution} can be approximated by
\begin{equation}\label{block-approximate}
    \left(\sum_{c=1}^{C}\mathbf{H}_{c}^{H}\mathbf{R}_{cc}^{-1}\mathbf{H}_{c}+\frac{1}{E_{s}}\mathbf{I}\right)^{-1}
    \left[
        \mathbf{H}_{1}^{H}\mathbf{R}_{11}^{-1},  \cdots  ,\mathbf{H}_{C}^{H}\mathbf{R}_{CC}^{-1}
    \right].
\end{equation}
Obviously, \eqref{block-approximate} can be implemented in a decentralized manner since each term in the summation can be computed locally. In the rest of this paper, the decentralized implementation of the above approximate equalization matrix \eqref{block-approximate} is called the block diagonal approximate correlation MMSE (BDAC-MMSE) algorithm. Although such an approximation will lead to a significant performance loss, it can still serve as a good initialization for our proposed decentralized method in the next.

\section{BCD-BASED MMSE EQUALIZATION}
In the decentralized daisy chain architecture, each DBU individually calculates its equalization matrix, in turn, using only local channel matrix $\mathbf{H}_{c}$, local noise samples $\mathbf{n}_{c}^{i}$, and the low dimensional updated information from the previous DBU. Finally, after several iterations, each DBU obtains a local equalization matrix.

Note that the equalization matrix $\mathbf{W}$ in \eqref{MMSE-model} can also be partitioned as $\mathbf{W}=[\mathbf{W}_{1},\mathbf{W}_{2},\cdots,\mathbf{W}_{C}]$. A popular approach to solve the block variable optimization problem is the BCD method \cite{2016Bertsekas}. At each iteration of this method, the function is minimized with respect to a single block of variables while the rest are held fixed. Specifically, the block variable $\mathbf{W}_{c}$ is updated by solving the following problem:
\begin{equation}\label{BCDMMSE-model}
    \min _{\mathbf{W}_{c}}  \quad \mathbb{E}\left\| \mathbf{W}_{c}\mathbf{y}_{c} +\sum_{j=1,j \neq c}^{C}\mathbf{W}_{j}\mathbf{y}_{j}-\mathbf{s} \right\|_2^{2}.
\end{equation}
Since the objective function of \eqref{BCDMMSE-model} is convex in $\mathbf{W}_{c}$, we could obtain the optimal solution by setting the gradient equal to $\mathbf{0}$, which yields
\begin{equation}\begin{aligned}\label{BCDMMSE-solution}
        \mathbf{W}_{c}^{\text{opt}}= & \left(E_s\left(\mathbf{I}_{K}-\sum_{j \neq c}\mathbf{W}_{j}\mathbf{H}_{j}\right)\mathbf{H}_{c}^{H}-\sum_{j \neq c}\mathbf{W}_{j}\mathbf{R}_{jc}\right) \\
                                     & \left(E_s\mathbf{H}_{c}\mathbf{H}_{c}^{H}+\mathbf{R}_{cc}\right)^{-1}.
    \end{aligned}\end{equation}
    
    \begin{algorithm}[!tb]
    \caption{BCD-based MMSE Equalization}
    \label{alg-BCD}
    Given the total iteration number $L.$\\
    \LinesNumbered 
    \KwIn{$\mathbf{H}=[\mathbf{H}_{1}^{T},\mathbf{H}_{2}^{T},\cdots,\mathbf{H}_{C}^{T}]^{T}, \mathbf{n}^i=[\mathbf{n}_1^i,\mathbf{n}_2^i,\cdots,\mathbf{n}_C^i],i=1,2,\cdots,N$ and $E_s$.}
    \textit{\textbf{Preprocessing:}}\\
    Initialize $\mathbf{W}^{0}$ using \eqref{block-approximate} in a decentralized manner\;
    $\mathbf{A}^{0} \gets \mathbf{0}$\;
    $\mathbf{b}_{i}^{0} \gets \mathbf{0},i=1,2,\cdots,N$\;
    \For{$c=1$ to $C$}{
        $\mathbf{A}^{0} \gets \mathbf{A}^{0}+\mathbf{W}^{0}_{c}\mathbf{H}_{c}$\;
        $\mathbf{b}_{i}^{0} \gets \mathbf{b}_{i}^{0}+\mathbf{W}^{0}_{c}\mathbf{n}_{c}^{i},i=1,2,\cdots,N$\;
    }
    \For{$c=1$ to $C-1$}{
        $\mathbf{A}^{0}_{c}\gets\mathbf{A}^{0}$\;
        $\mathbf{b}_{c,i}^{0}\gets\mathbf{b}_{i}^{0},i=1,2,\cdots,N$\;
    }
    \textit{\textbf{BCD iterations:}}\\
    \For{$l=1$ to $L$}{
        \For{$c=1$ to $C$}{
            Update $\mathbf{W}_{c}^{l}$ via \eqref{BCDMMSE-specific-update}\;
            Update $\mathbf{A}^{l}_{c}$ via \eqref{A-update}\;
            Update $\mathbf{b}_{c,i}^{l},i=1,2,\cdots,N$ via \eqref{b-update}\;
        }
    }
    \KwOut{$\mathbf{W}=[\mathbf{W}_{1},\mathbf{W}_{2},\cdots,\mathbf{W}_{C}]$.}
\end{algorithm}

Adopting the Gauss-Seidel update rule \cite{Hong2017BCD}, at $l$-th iteration, the block variable $\mathbf{W}_{c}^{l}$ is updated by solving the following subproblem:
\begin{equation}\label{BCDMMSE-update}
    \mathbf{W}_{c}^{l}=\arg\min _{\mathbf{W}_{c}} \quad f\left( \mathbf{W}_{1}^{l},\cdots,\mathbf{W}_{c-1}^{l},\mathbf{W}_{c},\mathbf{W}_{c+1}^{l-1},\cdots,\mathbf{W}_{C}^{l-1}\right).
\end{equation}
\begin{figure}[!tb]
    \centering
    \includegraphics[width=0.5\textwidth]{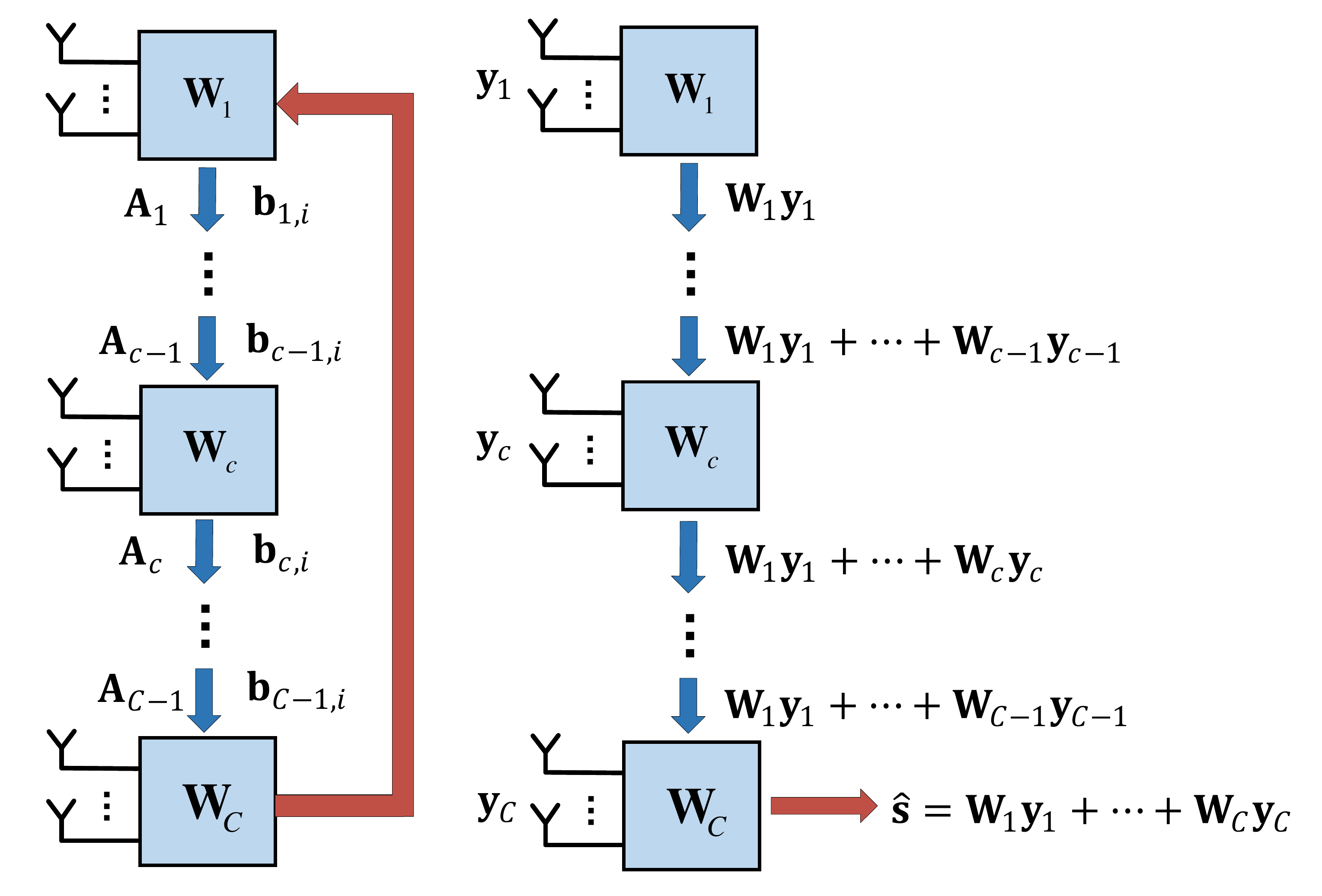}
    \caption{Illustration of information interconnection in proposed BCD-MMSE equalizer.}
    \label{fig:information}
\end{figure}
Based on \eqref{BCDMMSE-solution} and \eqref{BCDMMSE-update}, we  derive a  BCD-MMSE algorithm for the uplink equalization. Specifically, $\mathbf{W}_{c}^{l}$ is updated by:
\begin{dmath}\label{BCDMMSE-specific-update}
    \mathbf{W}_{c}^{l} = \left(E_s\left(\mathbf{I}_{K}-\mathbf{A}^{l}_{c-1}+\mathbf{W}^{l-1}_{c}\mathbf{H}_{c}\right)\mathbf{H}_{c}^{H} -\frac{1}{N}\sum_{i=1}^{N}\left(\left(\mathbf{b}_{c,i}^{l-1}-\mathbf{W}_{c}^{l-1}\mathbf{n}_{c}^{i}\right)(\mathbf{n}_{c}^{i})^{H}\right)\right) \\ \left(E_s\mathbf{H}_{c}\mathbf{H}_{c}^{H}+\hat{\mathbf{R}}_{cc}\right)^{-1}
\end{dmath}
where $\mathbf{A}^{l}_{c}$ and $\mathbf{b}_{c,i}^{l}$ is updated by the following two equations respectively.
\begin{equation}\label{A-update}
    \mathbf{A}^{l}_{c}=\mathbf{A}^{l}_{c-1}-\mathbf{W}^{l-1}_{c}\mathbf{H}_{c}+\mathbf{W}^{l}_{c}\mathbf{H}_{c},
\end{equation}
\begin{equation}\label{b-update}
    \mathbf{b}_{c,i}^{l}=\mathbf{b}_{c,i}^{l-1}-\mathbf{W}^{l-1}_{c}\mathbf{n}_{c}^{i}+\mathbf{W}^{l}_{c}\mathbf{n}_{c}^{i},i=1,2,\cdots,N.
\end{equation}

The BCD-based MMSE equalization is summarized in Algorithm \ref{alg-BCD}. The left part of Fig.~\ref{fig:information} shows the information interconnection in the proposed BCD-MMSE equalization matrix computation, where the superscript $l$ is omitted for brevity, the right part of Fig.~\ref{fig:information} shows the process of the equalizer filter to obtain the estimate of the transmitted symbol. Note that $\mathbf{W}_{c}^l\in \mathbb{C}^{K \times M_c}$ can be seen as a dimensionality reduction matrix to reduce the dimension of $\mathbf{H}_{c}$, $\mathbf{n}_{c}^{i}$, and $\mathbf{y}_{c}$ before interconnection. For example, to share the correlation matrix information among DBUs, the BCD-MMSE only needs exchange $\mathbf{b}^l_{c,i} \in \mathbb{C}^{K}$ between DBUs, which significantly reduce the amount of information interconnection when compared to directly exchange $\mathbf{n}_{c}^{i} \in \mathbb{C}^{M_c}$.
\begin{remark}
(Convergence Analysis) The convergence of Algorithm
\ref{alg-BCD} is guaranteed by the adopted BCD framework  \cite{2016Bertsekas}. More specifically, since the objective function of \eqref{BCDMMSE-model} is a continuously differentiable strongly convex function, after fixing all the optimization variables
other than one variable, the resulting subproblem is still a strongly convex optimization problem. Therefore, it was shown in [10, page 278]
that Algorithm 1 is guaranteed to converge to the global minimum of problem \eqref{BCDMMSE-model}.
\end{remark}
\begin{figure}[!tb]
    \centering
    \includegraphics[width=0.4\textwidth]{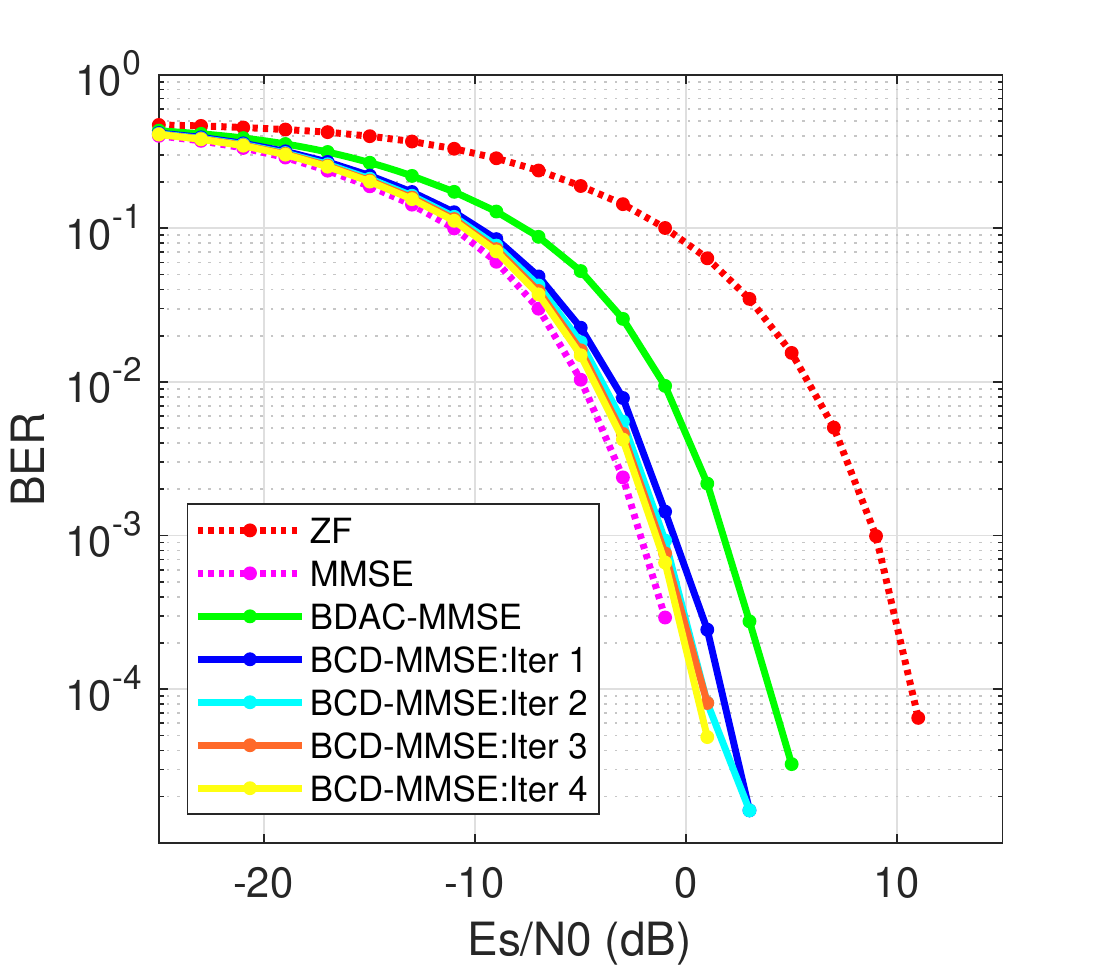}
    \caption{BER performance of first 4 iterations of BCD-MMSE with $\text{IoT} = 10$.}
    \label{fig:ber_iterations}
\end{figure}
The computational complexity is evaluated by counting the number of complex-valued matrix-matrix multiplication and matrix inversion operation. Since $N>M_c$ and $N>K$, the complexity of the initialization of the proposed BCD-MMSE, i.e. the BDAC-MMSE, is dominated by $\mathcal{O}(MK^2 + NMM_c)$, while the complexity of one iteration of the BCD-MMSE is $\mathcal{O}\left(NMK+MM_cK\right)$. On the other hand, the complexity of the centralized MMSE is $\mathcal{O}\left(M^3+NM^2\right)$, where $M$ may be extremely large in massive MIMO cases. Therefore, BCD-MMSE achieves far lower computational complexity by spreading the computing burden among multiple DBUs.


The number of transmitted complex-valued entries between any two adjacent DBUs during the BCD-MMSE algorithm is counted as $ (3K^2+2NK) + LK(N+K)$. The first part is caused by preprocessing, and the second part by $L$ iterations of the algorithm. Notably, the number of transmitted complex-valued entries is independent of the number of BS antennas $M$. Therefore, BCD-MMSE can achieve decentralized baseband processing design with a relatively small amount of data interconnection bandwidth among DBUs, enabling higher scalability and flexibility.

Based on the analysis above, the bottlenecks in terms of computation burden and data interconnection bandwidth can be mitigated to a large extent by BCD-MMSE.
\begin{remark}
The BCD-MMSE algorithm is proposed for the architecture in Fig.~\ref{fig_architecture}(a), but it can also be extended to other decentralized architectures by changing the update rule in \eqref{BCDMMSE-update}. For example, symmetric Gauss-Seidel update rule \cite{sun2016efficient} works well for the architecture in Fig.~\ref{fig_architecture}(b) and the convergence can be guaranteed. For the decentralized star architecture where a central DBU communicates with other DBUs, we could use the Jacobi update rule \cite{Bertsekas1997Distributed} for BCD iteration.
\end{remark}

\section{SIMULATION RESULTS}
\begin{figure*}[!tb]
    \centering
    \begin{tabular}{ccc}
        \includegraphics[width=0.3\textwidth]{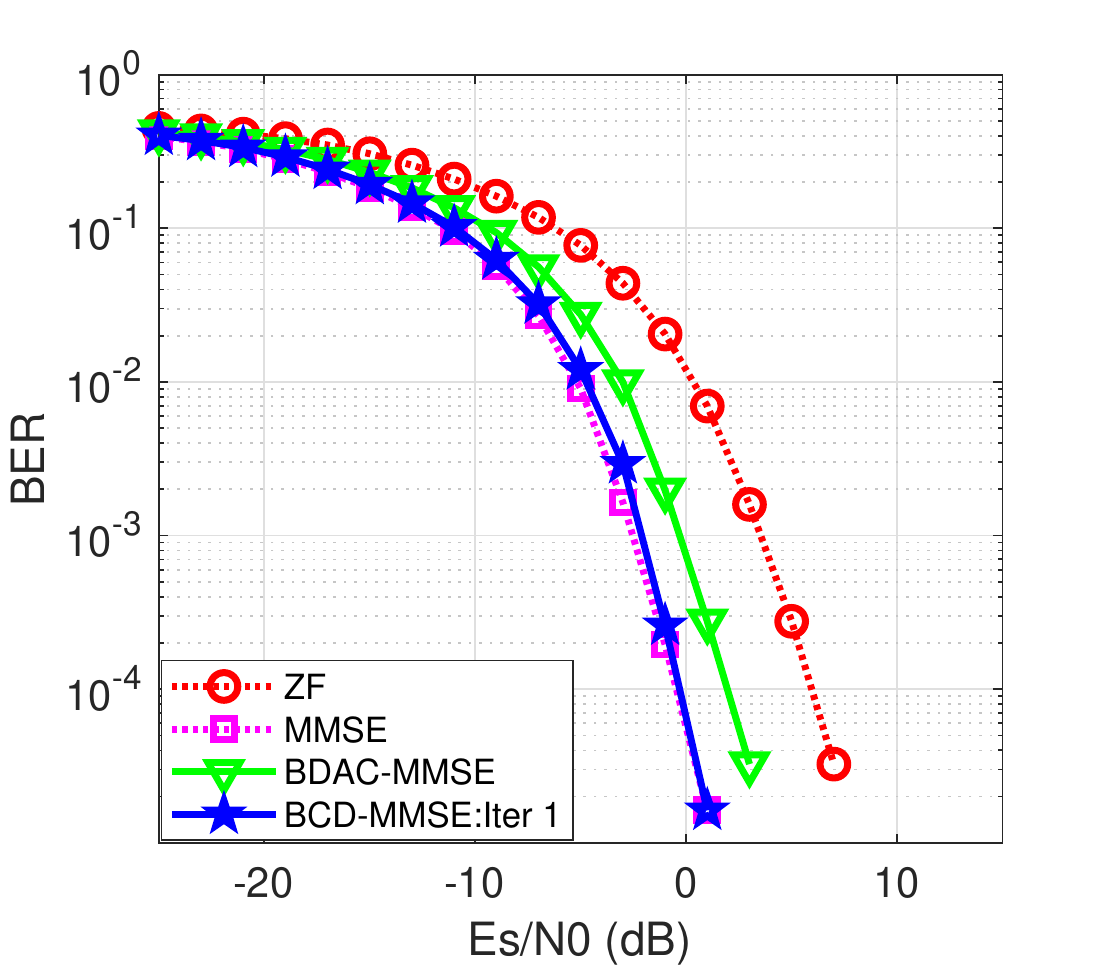} & \includegraphics[width=0.3\textwidth]{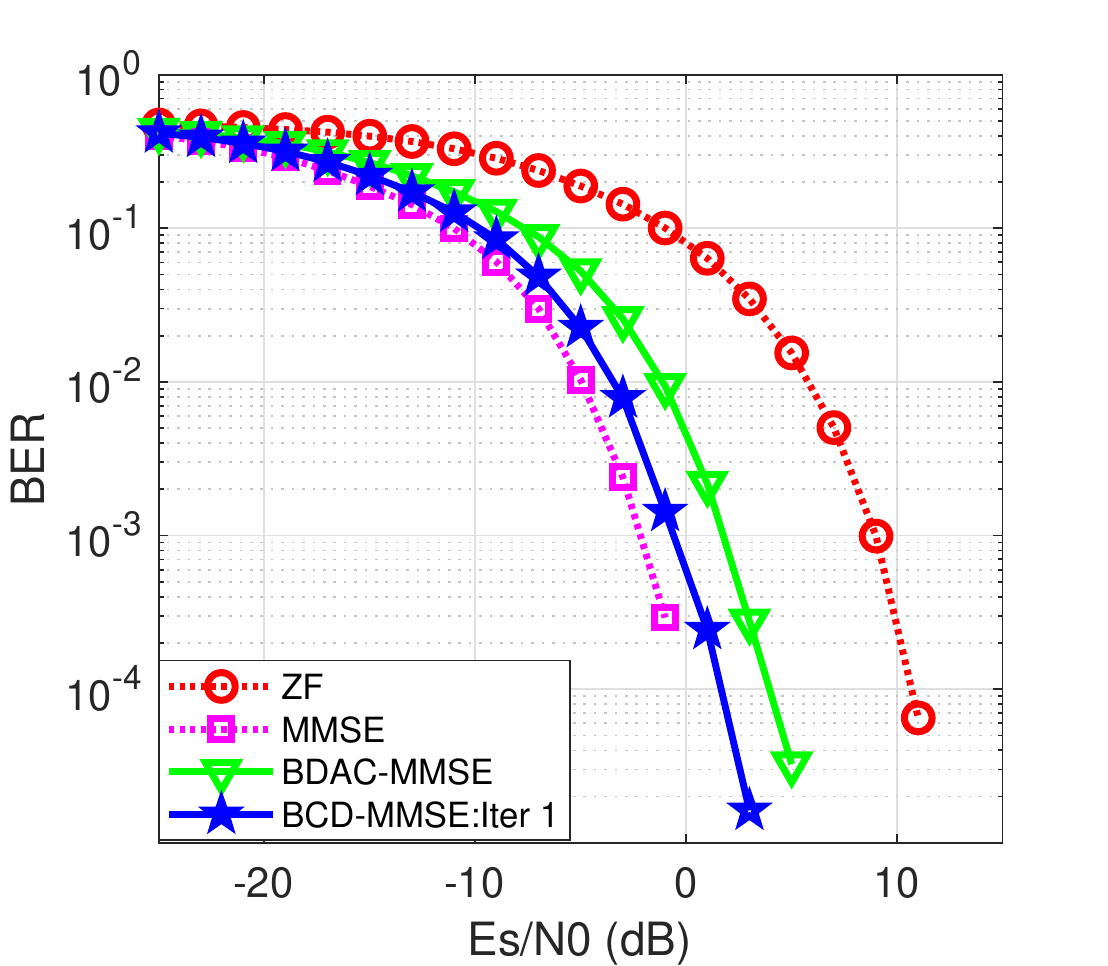} & \includegraphics[width=0.3\textwidth]{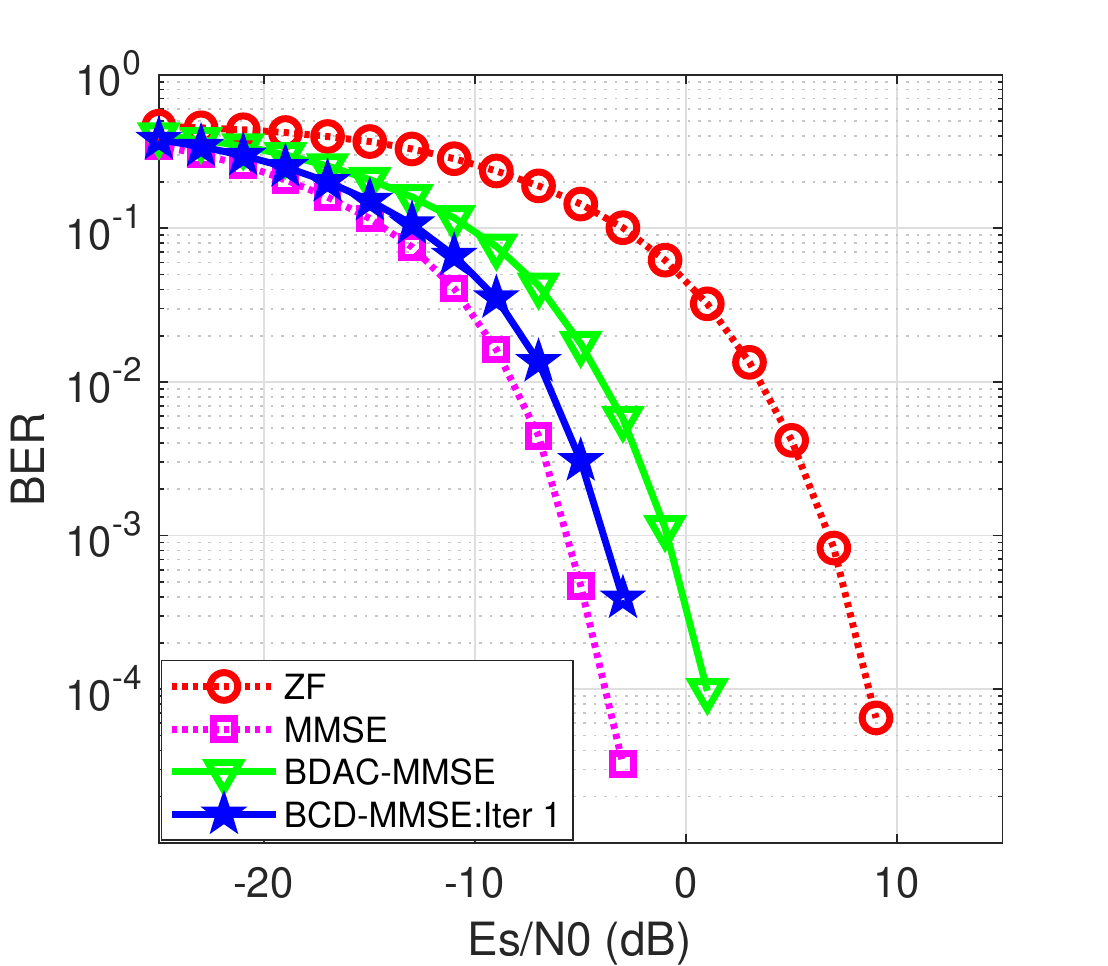} \\
        (a) IoT = 5, 128 antennas, 8 clusters. & (b) IoT = 10, 128 antennas, 8 clusters. & (c) IoT = 10, 256 antennas, 16 clusters.
    \end{tabular}
    \caption{BER performance under different intensities of interference and BS configurations}
    \label{fig:ber_iot}
\end{figure*}
In this section, the bit error rate (BER) performance of the proposed BCD-MMSE algorithm and other benchmarks are evaluated in an NR-based simulation system. Three baselines are considered in the simulation parts: including the commonly used centralized ZF method, the centralized MMSE that achieves near BER performance at the sacrifice of much more consumed computation and bandwidth resource, and the intuitive BDAC-MMSE algorithm, which acts as the initial point of the BCD-MMSE algorithm.

Consider a massive MIMO system consisting of a BS with $M=128$ antennas, which is equally divided into $C = 8$ clusters with each cluster size $M_c = 16$. The number of target UEs is $K = 8$  and the number of interference UEs is also 8. The number of noise samples equals $192$. The channel matrix is generated from QuaDRiGa platform \cite{quadriga} under the effects of large and small scale fading, where all the UEs are evenly scattered with an interval of 10 degrees on an arc centered at BS of radius $50$ to $100$.
We adopt $\text{Es/N0}$ to measure the normalized signal-to-noise ratio (SNR) and Interference over Thermal (IoT) to denote the interference to noise intensity ratio.
$\text{IoT} = 10$ is a default in the simulation since it is a typical value of a colored noise scene in practice. 

Fig.~\ref{fig:ber_iterations} shows the BER performance of the first four iterations of the BCD-MMSE algorithm, which illustrates the fast convergence property of the proposed BCD-MMSE.
Only 4 to 5 iterations of the BCD-MMSE equalization could approach the performance of centralized MMSE equalization. Even a single BCD iteration enables
excellent BER performance for data detection, enabling near-optimal performance while keeping required bandwidth and computational resources low.

The comparison of Fig.~\ref{fig:ber_iot}(a) and \ref{fig:ber_iot}(b) shows that the BCD-MMSE algorithm has excellent adaptability to different intensities of interference.
For a scenario with lighter IoT, it even performs equally well as centralized MMSE.
Fig.~\ref{fig:ber_iot}(c) reflects that the BCD-MMSE algorithm still has good performance in a more extensive antenna system, revealing its excellent scalability and the excellent potential for its application to extreme-large MIMO systems in the near future.

\section{CONCLUSIONS}
This paper has proposed a novel BCD-based MMSE equalization in a decentralized daisy chain architecture,
which mitigates both the computation and interconnection bottlenecks in typical centralized designs. Unlike previous works based on AWGN assumptions, we consider interference users in uplink communication systems, which lead the noise correlation matrix to be non-diagonal and challenging for decentralized implementation.
Extensive numerical results have shown that our approach can achieve a performance close to the centralized MMSE equalization for a large-scale system.

\end{document}